\documentclass[aps,amsmath,amssymb,twocolumn,10pt,pra,superscriptaddress]{revtex4-1}
 \usepackage{graphicx} 
 \usepackage[dvipsnames]{xcolor}
 \usepackage[colorlinks=true,linkcolor=blue,citecolor=red,urlcolor=magenta]{hyperref}
 \usepackage{charter}
 \usepackage{graphicx}
 \usepackage{dcolumn}
 \usepackage{bm}
 \usepackage{hyperref}
 \usepackage{amssymb}
 \usepackage{hyperref} 
 \usepackage{braket}
 \begin{document}

 \title{
  Radial and Angular Correlations in a Confined System of Two Atoms in Two-Dimensional Geometry
 }
 \author{Przemys{\l}aw Ko\'scik}
 \email{p\_koscik@pwsztar.edu.pl}
 \affiliation{Department of Computer Sciences, University of Applied Sciences, Mickiewicza 8, PL-33100 Tarn\'{o}w, Poland}
 \begin{abstract}

 We study the ground-state correlations between two atoms in a two-dimensional isotropic harmonic trap. We consider a finite-range soft-core interaction that can be applied to simulate various atomic systems. We provide detailed results on the dependence of the correlations on  the parameters of the system. Our investigations show that in  the hardcore limit, the wave function can be  approximated as the product of the radial and angular components regardless of the interaction range. This implies that the radial and angular correlations are independent of one another. However, correlations within the radial and angular components persist and are heavily influenced by the interaction range. The radial correlations are generally weaker than the angular correlations. When soft-core interactions are considered, the correlations exhibit more complex behavior.
 \end{abstract} 
 
 \maketitle
 \section{ INTRODUCTION} In recent years, there has been a growing interest in studying the properties of systems consisting of particles confined in external potentials \cite{ 0,1,2,2.1,3, oo,mig,4,5,5.1,6,7,8,9,10}. This interest has been fuelled by significant experimental advances that  enabled the creation of such systems in laboratories.
 Entanglement in these systems has received considerable attention because of its relevance to quantum information technology and the challenge of quantifying the degree of correlation. Significant scientific effort has been dedicated to understanding the correlation properties of harmonically trapped systems. These systems include those with harmonics \cite{mosh}, delta contact \cite{bospiv,boscor}, finite-range soft-core \cite{step}, inverse power law \cite{pl1,pl2,pl3,pl4,pl5}, and $\wp$-wave interactions \cite{10,bospiv}. Several studies have explored entanglement in natural systems such as helium and helium-like atoms \cite{hel1, hel2, hel3,hel4}. A review of  entanglement in composite systems, including atoms and molecules, can be found in \cite{rev}. In addition, recent advances in machine learning and deep learning are opening up new opportunities to study entanglement properties in various quantum systems \cite{NN, NNj}, including few-body systems \cite{NN1, NN2, NN3}\\
 
 The goal of our research is to better understand the correlation properties of a two-dimensional system consisting of  two bosonic atoms in a harmonic trap 
 that interact through a finite-range soft-core potential.\\

 The Hamiltonian that describes this system is given by
 \begin{equation}\label{Hamiltonian_total}
 {\cal H} = \sum_{i=1}^2\left[-\frac{\hbar^2\nabla^{2}_{\mathbf{r}_{i}}}{2\mathrm{m}} + V(\mathbf{r}_i)\right]+U(|\mathbf{r}_1-\mathbf{r}_2|),
 \end{equation}
 where $V(\mathbf{r})=m\omega^2 \mathbf{r}^2/2$ and 
 \begin{equation}\label{int}
 { U}(\bold{r}) = \left\{
 \begin{tabular}{lcl}
 $\kappa$, & $0\leq$&$\boldsymbol{r}\leq\sigma$ \\
 $0$, & $\boldsymbol{r}>\sigma,$
 \end{tabular}
 \right.\
 \end{equation}
 where
 $\kappa$ and $\sigma$ are the strength and range of interactions, respectively.
 Interaction (\ref{int}) can be applied to simulate various atomic systems \cite{sort1,sort2,sort3,sort4,sort5}, including those with Rydberg-dressed atoms \cite{rydberg1,rydberg2,rydberg3,rydberg4}.
 The analysis of this system can be simplified by separating the relative motion from the motion of the center of mass. 
 An additional advantage of this system is that it can be solved exactly \cite{ryd0,ryd1} using solutions expressed by special functions. Analytical closed-form solutions are available for certain control parameter values \cite{ryd1}.\\

Here, we present our findings in a structured manner. First, in Section \ref{Sec1}, we analyze particle correlations in the ground state across a broad range of control parameters. 
 In Section \ref{Sec2}, we examine the correlations between the subsystems associated with radial and angular variables. A notable observation in this section is that in the hardcore limit $\kappa\to \infty$, the correlations between the atoms can be accurately decomposed into radial and angular correlations. A detailed analysis of these correlations is provided. Finally, the conclusions in Section \ref{Sec3} highlight the main findings of this study.
  \section{PARTICLE CORRELATIONS}\label{Sec1}
 In this study, we focus on exploring the ground states of  two bosonic atoms. This state has a total orbital angular momentum equal to zero. It depends only on the radial coordinates $r_{1}$ and $r_{2}$ as well as the angular coordinate between the particles, $\theta_{12}=\varphi_{1}-\varphi_{2}$. $\Psi(\mathbf{r}_{1},\mathbf{r}_{2})={\Psi}(r_{1}, r_{2},\mathrm{cos}(\theta_{12}))$, where the normalization condition gives
 $2\pi\int_{0}^{2\pi}\int_{0}^{\infty}\int_{0}^{\infty} d\theta_{12}dr_{1}dr_{2}r_{1}r_{2}|\Psi|^2=1$. Schmidt decomposition is a tool for analyzing the correlations  between two bosons \cite{sch}. To obtain the Schmidt formula for the wave  function $\Psi$, we first decompose it into a Fourier-Lagrange series represented as follows:
 \begin{align}\label{fl}
 \sqrt{r_{1}r_{2}}\Psi(\mathbf{r}_{1},\mathbf{r}_{2})={g_{0}(r_{1}, r_{2})\over {2\pi}}+\sum_{l=1}{ g_{l}(r_{1},r_{2}) \mathrm{cos}[l(\varphi_{1}-\varphi_{2})]\over{\pi}},
 \end{align}
 where the term $\sqrt{r_{1}r_{2}}$ provides the correct normalization in the radial directions, and the component $g_{l}(r_{1}, r_{2})$ is calculated using the integral 
 \begin{align}\label{fur}
 g_{l}(r_{1}, r_{2})=\sqrt{r_{1}r_{2}}\int_{0}^{2\pi}d\theta_{12} \Psi(\mathbf{r}_{1},\mathbf{r}_{2}) \mathrm{cos}(l\theta_{12}).
 \end{align}
 The function $g_l(r_{1}, r_{2})$ is both real and symmetric, which means we can express it using the Schmidt decomposition as \begin{align}
 g_l(r_{1}, r_{2}) = \sum_{n=0} k_{nl} \chi_{nl}(r_1) \chi_{nl}(r_2),\end{align}
 where $\langle \chi_{nl} | \chi_{n'l} \rangle = \delta_{nn'}$. Using the above expression and the identity $\mathrm{cos}(l\theta) = (e^{\mathfrak{i} l\theta} + e^{-\mathfrak{i}l\theta})/2$ (where $\mathfrak{i}$ is an imaginary unit), we can represent the wave function $\Psi$ as
 \begin{align}\label{SD}
 \Psi(\mathbf{r}_{1},\mathbf{r}_{2})=\sum_{\substack{n=0\\l=-\infty...\infty}}k_{nl}u^{*}_{nl}(\mathbf{r}_{1})u_{nl}(\mathbf{r}_{2}),
 \end{align}
 where the single-particle orbitals are given by \begin{align}\label{Scm} 
 u^{}_{nl}(\mathbf{r})={\chi_{nl}(r)\over \sqrt{r
 }}{e^{\mathfrak{i} l\varphi}\over\sqrt{2\pi}},\end{align}
 $k_{nl}=k_{n|l|}$ and $\chi_{nl}(r)=\chi_{n|l|}(r)$. These orbitals form an orthonormal basis set; that is, $\int_{0}^{2\pi} \int_{0}^{\infty}d\varphi dr [r u^{*}_{nl}(\mathbf{r})u^{}_{n^{'}l^{'}}(\mathbf{r})]=\delta_{nn^{'}}\delta_{ll^{'}}$. Therefore, we conclude that Eq. (\ref{SD}) represents the Schmidt decomposition of the wave function $\Psi$. Note that both the orbital $u^{}_{nl}(\mathbf{r})$ and its complex conjugate are eigenfunctions of the angular momentum operator $(-\mathfrak{i}\hbar\partial_{\varphi})$ and the spatial reduced density matrix (RDM),
 \begin{align}
 \rho(\mathbf{r},\mathbf{r}^{'})=\int \Psi^{*}(\mathbf{r},\mathbf{r}_{2})\Psi(\mathbf{r}^{'},\mathbf{r}_{2})d\mathbf{r}_{2},
 \end{align}
 that is,
 \begin{align}
 \rho(\mathbf{r},\mathbf{r}^{'})=\sum_{\substack{n=0\\l=-\infty...\infty}}\lambda_{nl}u_{nl}(\mathbf{r})u^{*}_{nl}(\mathbf{r}^{'}),
 \end{align}
 where the eigenvalues $\lambda_{nl}$ (occupancies) are related to the Schmidt coefficients $k_{nl}$ by
 $\lambda_{nl}=k_{nl}^2$, all occupancies except those with $l=0$ are doubly degenerate and the conservation of probability yields $\sum_{nl=0}\lambda_{nl}+2\sum_{nl=1}\lambda_{nl}=1$.
 The state of two bosonic atoms is nonentangled if and only if it can be represented by a single permanent \cite{sch}. This occurs when
 there exists a Schmidt coefficient equal to either $k_{n0}=1$ or $k_{nl}=1/\sqrt{2}$.
 Note that the remaining Schmidt coefficients disappear in each of these cases owing to the conservation of  probability. 
\begin{figure}
 \includegraphics[width=87.2mm, height=87.2mm]{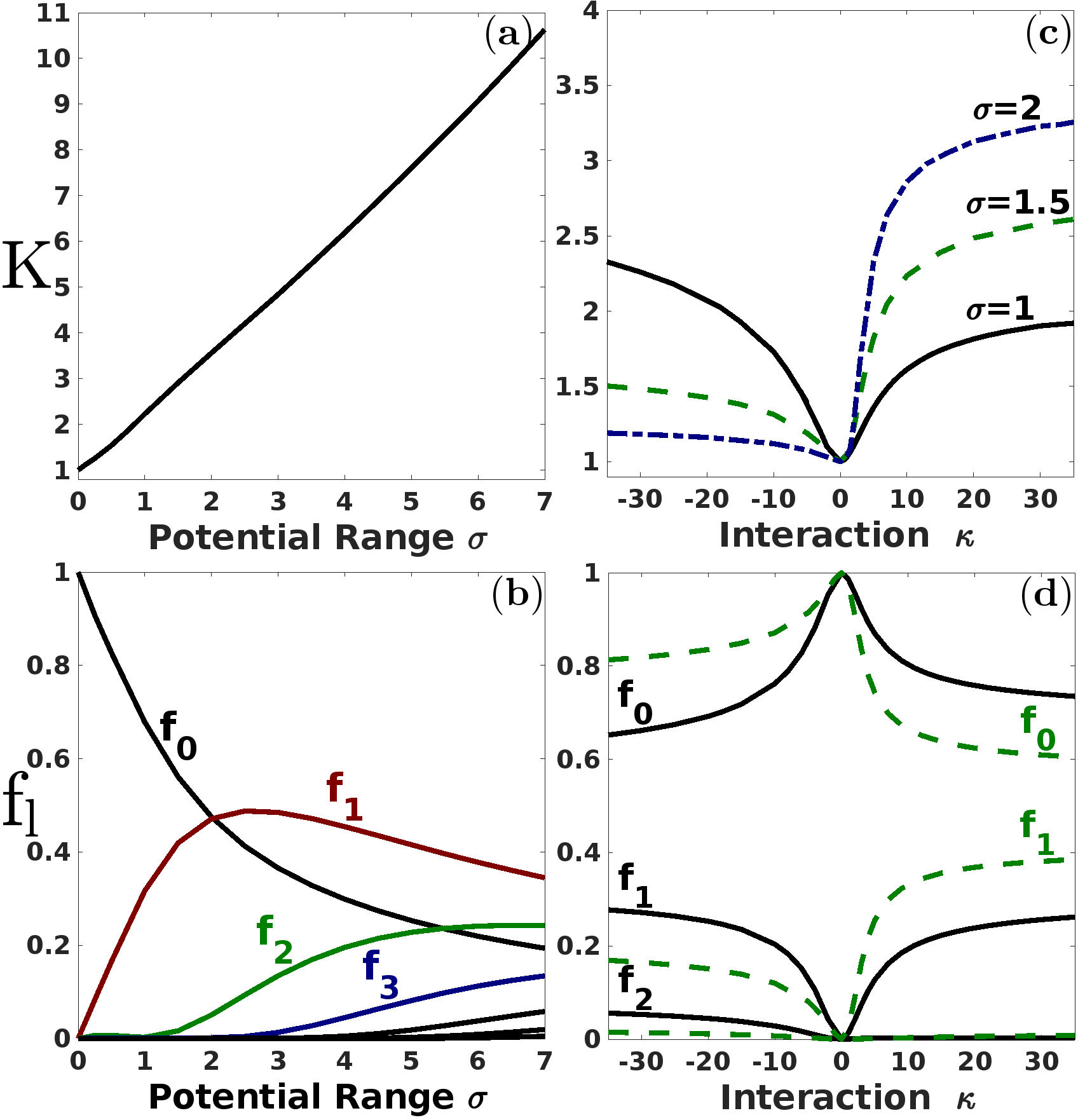}
 \caption{ The graphs labeled ({\bf a}) and ({\bf b}) depict the participation ratio $K$ and the  corresponding collective occupancies $f_{l}$  in the hardcore limit $\kappa\to\infty$ as functions of $\sigma$, respectively.
 Graph ({\bf c}) illustrates
 the participation ratio $K$ for the three different values of $\sigma$ as a function of $\kappa$. Plot ({\bf d}) 
 shows the collective occupancies $f_{l}$ 
 for $\sigma$ equal to $1$ (solid lines) and $1.5$ (dashed lines) as functions of $\kappa$. The interaction range $\sigma$ and strength $\kappa$ are  measured in $\sqrt{\hbar/\mathrm{m}\omega}$ and $\hbar\omega$, respectively.\label {Fig1} }
 \end{figure}

 To quantify the degree of correlation in the ground state $\Psi$, we use the participation ratio ${K}={\cal P}^{-1}$ \cite{part}, where ${\cal P}=\mathrm{Tr}\hat{\rho}^{2}$ is the purity of the RDM, which is given by \begin{align}{\cal P}=\sum_{{\substack{n=0\\l=-\infty...
\infty}}}\lambda_{nl}^2.\end{align}The participation $K$ counts approximately 
 the number of one-particle orbitals actively involved in Schmidt decomposition (\ref{SD}). 
 A helpful tool for analyzing particle correlations  is the collective occupancy,   \begin{align}f_{l}=(2-{\delta}_{l0})\sum_{n}\lambda_{nl},\end{align}  which calculates the probability of finding a pair of particles, where one has an angular momentum of $\hbar l$ and the other has an angular momentum of $-\hbar l$, $\sum_{l=0}f_{l}=1$.
Note that the values of
 ${\cal P}$ and $f_{l}$ can be determined without explicitly calculating occupancies. Instead, they can be expressed in terms of integrals
 ${\cal P}=\sum_{l}\int[\rho_{l}(r,r^{'})]^2dr dr^{'}$ 
 and 
 $f_{l}=(2-{\delta}_{l0})\int \rho_{l}(r, r)dr$,
 where
 $\rho_{l}(r,r^{'})=\int g_{l}(r, r_{2})g_{l}(r^{'},r_{2})dr_{2}$.\\ 

Fig. \ref{Fig1} summarizes our results for particle correlations in the ground state.
 The left panel shows the results obtained for the hardcore limit $\kappa\to\infty$ as a function of $\sigma$. The plots labeled ({\bf a}) and ({\bf b}) depict the results of the participation ratio and corresponding collective occupancies, respectively. The participation ratio increases almost linearly with increasing $\sigma$, indicating that the number of significant Schmidt orbitals in Eq. (\ref{SD}) increases in a similar manner. The fraction of particle pairs with only radial correlations $(f_{0})$ decreases steadily with $\sigma$. In contrast, the collective occupancies for higher values of $l$ exhibit more complicated behavior. Initially, the value of collective occupancy $f_{1}$ increases with increasing $\sigma$ and reaches parity with $f_{0}$ around $\sigma= 2$. Beyond this point, the contribution of $f_{1}$ decreases, and the components with higher $l$ become more prominent, indicating more complex correlation effects.
 The right panel of Fig. \ref{Fig1} shows the results obtained for finite interaction strengths $\kappa$. We observe that the effect of varying $\sigma$ on the correlations in the attraction regime is the opposite of that in the repulsion regime. Notably, it is worth noting that increasing $\sigma$ leads to a significant expansion in the range of negative $\kappa$ values, where the entanglement is weak. The considered state becomes nearly unentangled ($K\approx 1$) over a wide range of attractive forces starting from $\sigma=2$.
 \section{RADIAL AND ANGULAR CORRELATIONS}\label{Sec2} Schmidt's formula for analyzing the correlations between subsystems associated with the radial variable $\vec{r}=(r_{1},r_{2})$ and angular variable $\vec{\varphi}=(\varphi_{1},\varphi_{2})$ is as follows: 
 \begin{align}\label{sd1}
 \sqrt{r_{1}r_{2}}\Psi(\mathbf{r}_{1},\mathbf{r}_{2})=\sum_{n}{q}_{n}{\cal V}_{n}(\vec{r}){\Phi}_{n}(\vec{\varphi}),
 \end{align}
 $\langle {\cal V}_{n}|{\cal V}_{n^{'}}\rangle=\langle {\Phi}_{n}|{\Phi }_{n^{'}}\rangle=\delta_{nn^{'}}$. To obtain form (\ref{sd1}), we proceed as follows. First, we note that the wave function $\Psi$ can be decomposed as 
 \begin{align}\label{sd2}
 \sqrt{r_{1}r_{2}}\Psi(\mathbf{r}_{1},\mathbf{r}_{2})=\sum_{lm}W_{lm}{\Theta}_{l}(\vec{\varphi}){ v}_{m}(\vec{r}),
 \end{align}
 where
 \begin{align}
 \Theta_{0}(\vec{\varphi})={1\over 2\pi},\Theta_{l}(\vec{\varphi})={\mathrm{cos}[l (\varphi_{1}-\varphi_{2})]\over \sqrt{2}\pi},\end{align}
 $\langle \Theta_{l}|\Theta_{l^{'}}\rangle=\delta_{ll^{'}}$
 and $\{{v}_{m}(\vec{r})\}$ is some set of orthonormal basis functions, $\langle v_{m}|v_{m^{'}}
 \rangle=\delta_{mm^{'}}$. The expansion coefficients are given by 
 \begin{align} W_{lm}=\sqrt{2-\delta_{l0}}\int g_{l}(r_{1},r_{2}){v}_{m}(\vec{r})dr_{1}dr_{2}. \end{align}
 Next, using the singular value decomposition theorem $\textbf{W}=\textbf{U}\mathbf{Q}\textbf{V}^T$, $\textbf{W}=[W_{lm}]$, we obtain the Schmidt coefficients ${q}_{n}=\mathbf{Q}_{nn}$ and the Schmidt orbitals
 \begin{align}{\cal V}_{n}(\vec{r})=\sum_{j}(\mathbf{V})_{jn}{v}_{j}(\vec{r}),\end{align}
 and
 \begin{align}
 \Phi_{n}(\vec{\varphi})=\sum_{j} (\textbf{U})_{jn}\Theta_{j}(\vec{\varphi}).\end{align}
 In our numerical calculations, we used a basis set $\{v_{m}\}$ with permanent ${v}_{m}(\vec{r})=\mathrm{per}[\tilde{v}_{m_{1}}(r_{1}),\tilde{v}_{m_{2}}(r_{2})]$ formed by single-particle orbitals $\tilde{v}_{s}(r)=(\sqrt{2/L})\mathrm{sin}(s \pi r/L)$ for a sufficiently large box size $L$. 
The eigenvalues of the RDMs for the radial and angular subsystems, denoted by ${\gamma}_{n}$, are identical and given by 
 ${\gamma}_{n}={q}_{n}^2$ ($\sum \gamma_{n}=1$).
 To assess the coupling between the radial and angular correlations, we rely on the largest eigenvalue 
 $\gamma_{0}$. A given state is the ideal product of the radial and angular components when ${\gamma}_{0}=1$, which is the case for the ground state without interactions. Alternatively, ${\gamma}_{0}$ can be interpreted as the fraction of a pair of particles with independent radial and angular correlations.
 \begin{figure} \includegraphics[width=75.8mm, height=65.8mm]{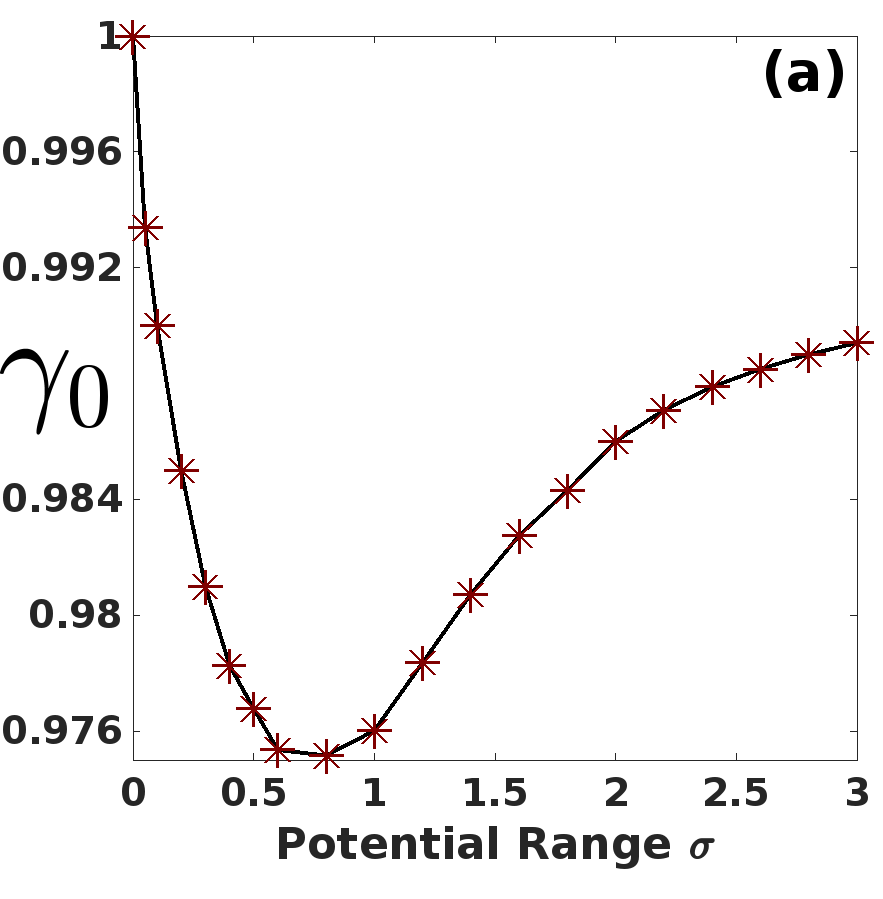} \includegraphics[width=86.2mm,height=40.8mm]{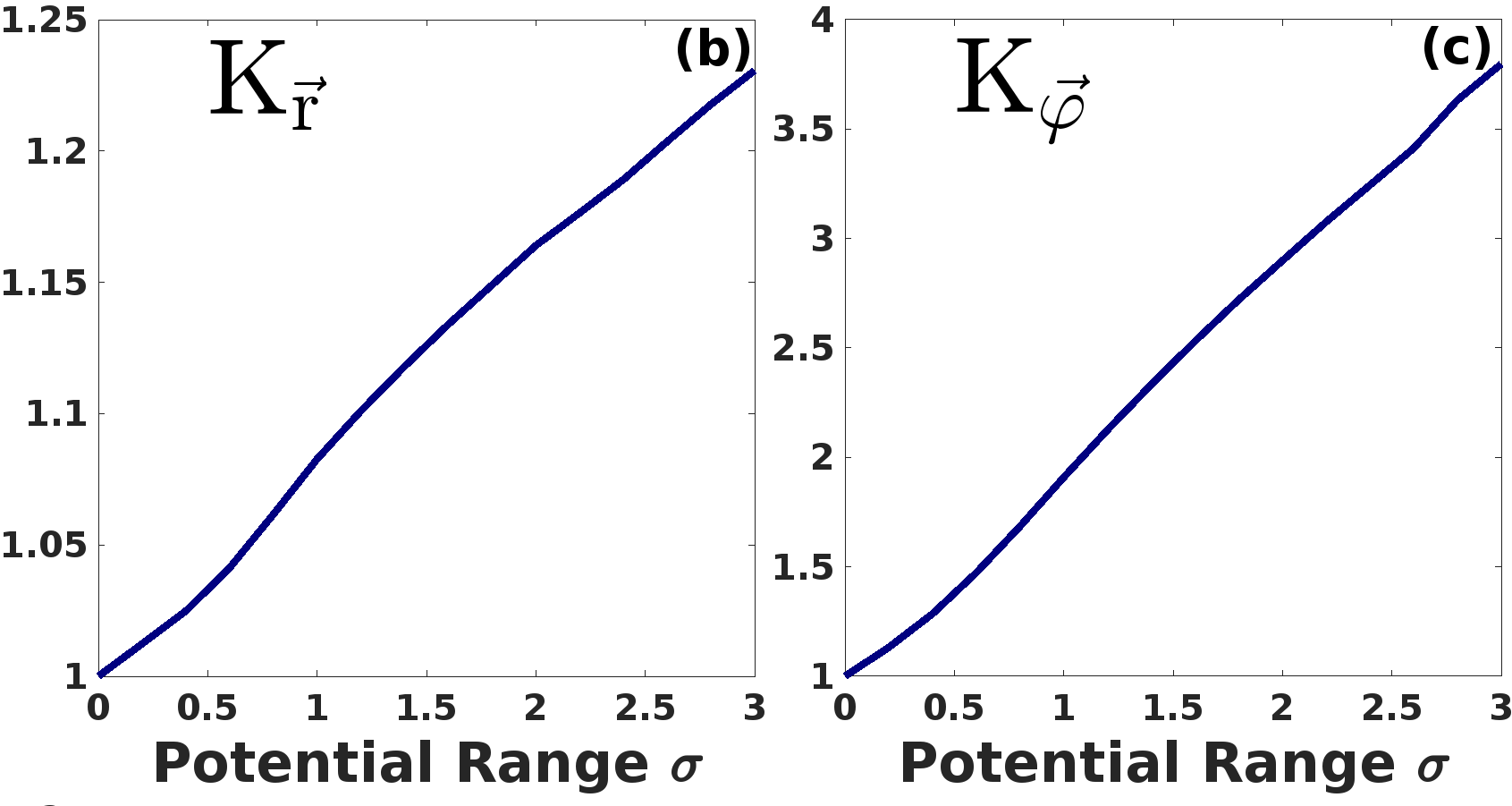} \caption{ Results obtained in the hardcore limit $\kappa\to\infty$ as a function of $\sigma$. Graph ({\bf a}) shows the behavior of  the largest eigenvalue of the RDM for the radial/angular subsystem ${\gamma}_{0}={q}_{0}^2$. Plots ({\bf b}) and ({\bf c}) show the behavior of the participation ratios  $K_{\vec{r}}$ and $K_{\vec{\varphi}}$, respectively. The potential range $\sigma$ is measured in $\sqrt{\hbar/\mathrm{m}\omega}$.\label{Fig2} }
 \end{figure}
  First, we focus on the limit $\kappa\to \infty$. Our analysis indicates that ${\gamma}_{0}$ is almost equal to one regardless of the interaction range $\sigma$. This  is illustrated in Fig. \ref{Fig2} (\textbf{a}), wherein we also note a noteworthy  feature minimum at  approximately $\sigma= 0.8$. Based on this finding, we can conclude that the following approximation is valid: \begin{align}\label{prod1}
 \sqrt{r_{1}r_{2}}\Psi(\mathbf{r}_{1},\mathbf{r}_{2})\approx {\cal V}_{0}(r_{1},r_{2}){\Phi}_{0}(\varphi_{1},\varphi_{2}).\end{align} This means that the correlations between the radial and angular coordinates are nearly negligible.
 However, correlations still exist within both radial and angular variables. This is shown in Fig. \ref{Fig2} ({\bf b}) and Fig. \ref{Fig2} ({\bf c}), where $K_{\vec{r}}$ and $K_{\vec{\varphi}}$ represent the participation ratios calculated for the angular and radial components ${\cal V}_{0}(r_{1},r_{2})$ and ${\Phi}_{0}(\varphi_{1},\varphi_{2})$, respectively.
 As shown, the radial correlations are generally weaker than the angular correlations, and the difference becomes more pronounced as $\sigma$ increases. 
 To obtain the Schmidt form of ${\cal V}_{0}(r_{1},r_{2})$, diagonalization is necessary, whereas the function ${\Phi}_{0}(\varphi_{1},\varphi_{2})$ can be automatically written in Schmidt form as follows:
 \begin{align}\Phi_{0}({\varphi}_{1},\varphi_{2})=\sum_{l=-\infty...\infty}w_{l} \phi^{*}_{l}(\varphi_{1})\phi_{l}(\varphi_{2}),\end{align}
 where $\phi_{l}(\varphi)={e^{\mathfrak{i} l\varphi}/\sqrt{2\pi}} $, $ w_{l}=(\textbf{U})_{|l|0}/\sqrt{2-\delta_{l0}}$. The participation ratio expressed by $w_{l}$ is  $K_{\vec{\varphi}}=\sum_{l} w_{l}^4$.\\

\begin{figure} \includegraphics[width=80.8mm, height=66.8mm]{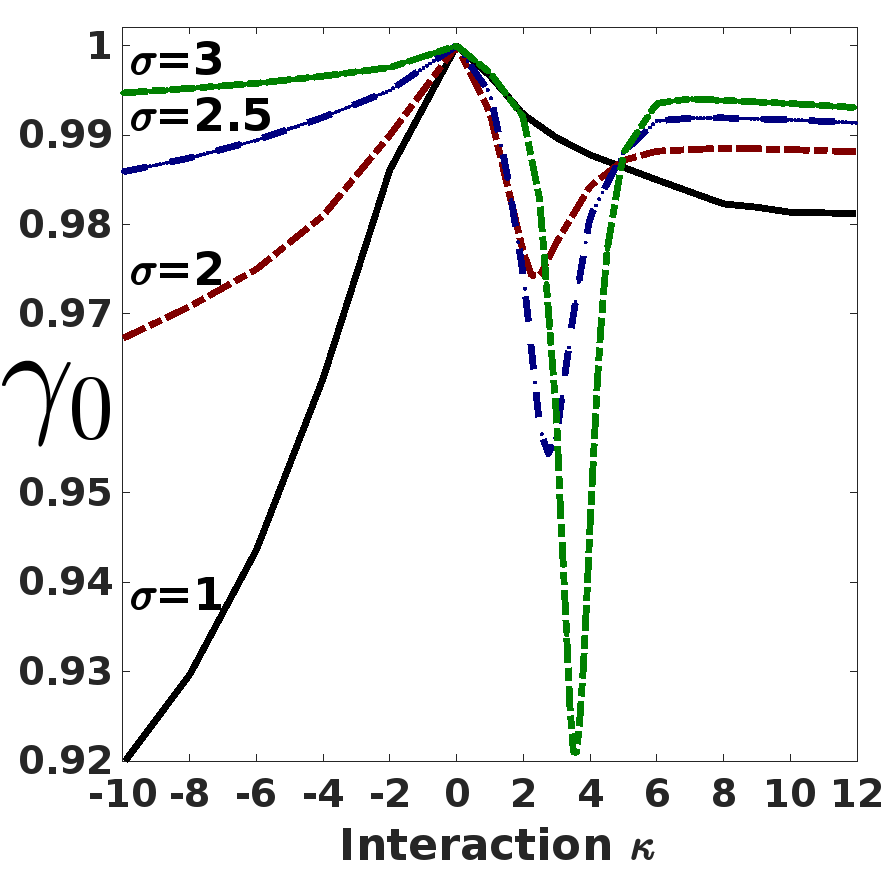}
 \caption{ The plot shows the 
behavior of the eigenvalue $\gamma_{0}$ in
a soft-core interaction scenario. The  range $\sigma$ and strength $\kappa$ are  measured in $\sqrt{\hbar/\mathrm{m}\omega}$ and $\hbar\omega$, respectively.
  \label{Fig3} }
 \end{figure}
 
The distributions $n(r_{1},r_{2})=\int r_{1}r_{2}|\Psi|^2d\vec{\varphi}$ and $\Gamma(\varphi_{1},\varphi_{2})=\int r_{1}r_{2}|\Psi|^2d\vec{r}$
 encode spatial radial and angular  correlations. Sample results  regarding their   behaviors  can be found in \cite{ryd1}.
 Based on our current results, we conclude that the universal feature is that in the strong repulsion limit,  $n(r_{1},r_{2})\approx |{\cal V}_{0}(r_{1},r_{2})|^2$ and $\Gamma(\varphi_{1},\varphi_{2})\approx|\Phi_{0}(\varphi_{1},\varphi_{2})|^2$.
 In this limit, the radial relative motion wave function in terms of $r=|\mathbf{r}_{1}-\mathbf{r}_{2}|$ exhibits a maximum at a value of $r$ greater than $\sigma$. For a sufficiently large value of $\sigma$, this maximum becomes well localized, and a clear maximum in the behavior of  $n(\varphi_{1},\varphi_{2})$ appears at $\theta_{12}\approx \pi$. This process indicates that the system is undergoing crystallization.      Increasing $\sigma$ improves the localization of atoms on opposite sides of the trap.\\

For completeness, we also computed the value of $\gamma_{0}$ for finite values of $\kappa$. Our research has shown that for specific positive values of $\kappa$ and $\sigma$, the approximation in Eq. (\ref{prod1}) may not be as accurate as when $\kappa\to\infty$. This is illustrated in Fig. \ref{Fig3}, where it can be seen that when $\sigma$ is sufficiently large, $\gamma_{0}$ exhibits a local minimum. The position of this minimum shifts toward larger values of $\kappa$, and its value decreases as $\sigma$ increases.
 Interestingly, we found that the minimum in $\gamma_{0}$ is accompanied by the appearance of a maximum in the relative motion wave function at approximately $r={\sigma}$. The parameter values at which the minimum occurred can be interpreted as the transition point to a regime where crystallization is allowed with a significant probability.
 Regarding the  attractive forces, we observe that for fixed $\sigma$, $\gamma_{0}$ decreases monotonically as 
 $\kappa$ decreases. 
\section{CONCLUSIONS}\label{Sec3}We studied the correlation between two atoms interacting through a finite-range soft-core potential and confined in a harmonic trap. By employing Schmidt decomposition, we  determined the correlations between the particles and between the subsystems associated with their radial and angular variables.
When the interaction range is fixed, the entanglement between the particles is a monotonic function in both the repulsive and attractive regimes. Our results showed that the coupling between the radial and angular correlations has a complicated dependence on the system parameters. However, in the hardcore limit, it becomes almost insignificant and  independent of the interaction range.
  Consequently, in this limit, the particle correlations are almost exactly decomposed into radial and angular correlations, which can be analyzed separately.  
 Entanglement in both radial and angular components increases
monotonically as the interaction range increases.\\

 Our results enhance the understanding of quantum correlations and suggest that studying the correlations between the radial and angular subsystems can yield further insights into particle correlations. To gain a deeper understanding, it is crucial to examine how factors such as  the type of interaction, particle number, and confinement potential affect these correlations.

 \end{document}